\documentclass[twoside,11pt]{article}
\usepackage[usenames,dvipsnames]{color}

\usepackage{jmlr2e}

\usepackage{amsmath}

\usepackage{graphicx}
\usepackage{graphics}
\usepackage{epstopdf}
\usepackage{tabularx, multirow, booktabs, footnote}

\newtheorem{assumption}{Condition}

\newcommand{\pr}{\mathrm{pr}}

\newcommand{\argmin}{\mathrm{argmin}}

\newcommand{\bb}{\mbox{\bf b}}

\newcommand{\bh}{\mbox{\bf h}}

\newcommand{\bv}{\mbox{\bf v}}

\newcommand{\bx}{\mbox{\bf x}}
\newcommand{\by}{\mbox{\bf y}}

\newcommand{\bX}{\mbox{\bf X}}

\newcommand{\bzero}{\mbox{\bf 0}}
\newcommand{\bveps}{\mbox{\boldmath $\varepsilon$}}

\newcommand{\bbeta}{\mbox{\boldmath $\beta$}}
\newcommand{\bdelta}{\mbox{\boldmath $\delta$}}

\newcommand{\bGamma}{\mbox{\boldmath $\Gamma$}}

\newcommand{\bSig}{\mbox{\boldmath $\Sigma$}}

\newcommand{\hbbeta}{\widehat\bbeta}

\newcommand{\hbeta}{\widehat{\beta}}

\newcommand{\sgn}{\mathrm{sgn}}
\newcommand{\supp}{\mathrm{supp}}

\newcommand{\FS}{\text{FS}}

\def\t{^T}

\jmlrheading{?}{2015}{1--22}{12/14}{?}{Yinfei Kong, Zemin Zheng and Jinchi Lv}

\ShortHeadings{Constrained Dantzig Selector}{Kong, Zheng and Lv}
\firstpageno{1}

\begin{document}

\title{
The Constrained Dantzig Selector with Enhanced Consistency
\thanks{This work was supported by the NSF CAREER Award DMS-0955316, a grant from the Simons Foundation, and USC Diploma in Innovation Grant. We sincerely thank the Action Editor and two referees for their valuable comments that have helped improve the paper significantly.}}

\author{\name Yinfei Kong \email yinfeiko@usc.edu \\
       \addr Department of Preventive Medicine\\
       University of Southern California\\
       Los Angeles, CA 90089, USA
       \AND
       \name Zemin Zheng {\email zhengzm@ustc.edu.cn} \\
       \addr Department of Statistics and Finance\\
       University of Science and Technololgy of China\\
       Hefei, Anhui 230026, China
       \AND
       \name Jinchi Lv \email jinchilv@marshall.usc.edu \\
       \addr Data Sciences and Operations Department\\
       Marshall School of Business\\
       University of Southern California\\
       Los Angeles, CA 90089, USA
       }


\editor{Sara van de Geer}

\maketitle

\begin{abstract}
The Dantzig selector has received popularity for many applications such as compressed sensing and sparse modeling, thanks to its computational efficiency as a linear programming problem and its nice sampling properties. Existing results show that it can recover sparse signals mimicking the accuracy of the ideal procedure, up to a logarithmic factor of the dimensionality. Such a factor has been shown to hold for many regularization methods. An important question is whether this factor can be reduced to a logarithmic factor of the sample size in ultra-high dimensions under mild regularity conditions. 
To provide an affirmative answer, in this paper we suggest the constrained Dantzig selector, which has more flexible constraints and parameter space.
We prove that the suggested method can achieve convergence rates within a logarithmic factor of the sample size of the oracle rates and improved sparsity, under a fairly weak assumption on the signal strength. Such improvement is significant in ultra-high dimensions. This method can be implemented efficiently through sequential linear programming. Numerical studies confirm that the sample size needed for a certain level of accuracy in these problems can be much reduced.\\
\end{abstract}


\section{Introduction} \label{Sec1}

Due to the advances of technologies, big data problems appear increasingly common in the domains of molecular biology, machine learning, and economics. It is appealing to design procedures that can provide a recovery of important signals, among a pool of potentially huge number of signals, to a desired level of accuracy. As a powerful tool for producing interpretable models, sparse modeling via regularization has gained popularity for analyzing large-scale data sets. A common feature of the theories for many regularization methods is that the rates of convergence under the prediction or estimation loss usually involve the logarithmic factor of the dimensionality $\log p$; see, for example, Bickel et al. (2009), van de Geer et al. (2011), and Fan and Lv (2013), among many others. From the asymptotic point of view, such a factor may be negligible or insignificant when $p$ is not too large. It can, however, become no longer negligible or even significant in finite samples with large dimensionality, particularly when a relatively small sample size is considered or preferred. In such cases, the recovered signals and sparse models may tend to be noisy. Another consequence of this effect is that many noise variables tend to appear together with recovered important ones (Cand\`{e}s et al., 2008). An important and interesting question is whether such a factor can be reduced, say, to a logarithmic factor of the sample size $\log n$, from ultra-high dimensions under mild regularity conditions.

In high-dimensional variable selection, there is a fast growing literature for different kinds of regularization methods with well established rates of convergence under the estimation and prediction losses. For instance, Cand\`{e}s and Tao (2007) proposed the $L_1$-regularization approach of the Dantzig selector by relaxing the normal equation to allow for correlation between the residual vector and all the variables, which can recover sparse signals as accurately as the ideal procedure, knowing the true underlying sparse model, up to a logarithmic factor $\log p$. Later, Bickel et al. (2009) showed that the popularly used Lasso estimator (Tibshirani, 1996) exhibits similar behavior as the Dantzig selector with a $\log p$ factor in the oracle inequalities for the prediction risk and bounds on the estimation losses in general nonparametric regression models. Furthermore, it has been proved in Raskutti et al. (2011) that the minimax rates of convergence for estimating the true coefficient vector in the high-dimensional linear regression model involve a logarithmic factor $\log p$ in the $L_2$-loss and prediction loss, under some regularity conditions on the design matrix. Other work for desired properties such as the oracle property on the $L_1$ and more general regularization includes Fan and Li (2001), Fan and Peng (2004), Zou and Hastie (2005), Zou (2006), van de Geer (2008), Lv and Fan (2009), Antoniadis et al. (2010), St\"{a}dler et al. (2010), Fan and Lv (2011), Negahban et al. (2012), Chen et al. (2013), and Fan and Tang (2013), among many others.

A typical way for reducing the logarithmic factor $\log p$ to $\log n$ is through model selection consistency, where the estimator selects exactly the support of the true coefficient vector, that is, the set of variables with nonzero coefficients. We refer the reader to Zhao and Yu (2006), Wainwright (2009), Zhang (2010), and Zhang (2011) for analysis of model selection consistency of regularization methods. Since the true parameters are assumed to be sparse in the high-dimensional setting, consistent variable selection can greatly lessen the analytical complexity from large dimensionality $p$ to around oracle model size. Once model selection consistency is established for the estimator with significant probability, an analysis constrained on that event will give a factor $\log n$ instead of $\log p$ in the rates of convergence under various estimation and prediction losses. However, to obtain model selection consistency it usually requires a uniform signal strength condition that the minimum magnitude of nonzero coefficients is at least of order $\{s(\log p)/n\}^{1/2}$ (Fan and Lv, 2013 and Zheng et al., 2014), where $s$ stands for the size of the low-dimensional parameter vector. We doubt the necessity of the model selection consistency and the uniform signal strength of order $\{s(\log p)/n\}^{1/2}$ for achieving the logarithmic factor $\log n$ in ultra-high dimensionality.


In this paper, we suggest the constrained Dantzig selector to study the rates of convergence with weaker signal strength assumption. The constrained Dantzig selector replaces the constant Dantzig constraint on correlations between the variables and the residual vector by a more flexible one, and considers a constrained parameter space distinguishing between zero parameters and significantly nonzero parameters. The main contributions of this paper are threefold. First, the convergence rates for the constrained Dantzig selector are shown to be within a factor of $\log n$ of the oracle rates instead of $\log p$, a significant improvement in the case of ultra-high dimensionality and relatively small sample size. It is appealing that such an improvement is made with a fairly weak assumption on the signal strength without requiring model selection consistency. To the best of our knowledge, this assumption seems to be the weakest one in the literature of similar results; see, for example, Bickel et al. (2009) and Zheng et al. (2014). Two parallel theorems, under the uniform uncertainty principle condition and the restricted eigenvalue assumption, are established on the properties of the constrained Dantzig selector for compressed sensing and sparse modeling, respectively. Second, compared to the Dantzig selector, theoretical results of this paper show that the number of falsely discovered signs of our new selector, with an explicit inverse relationship to the signal strength, is controlled as a possibly asymptotically vanishing fraction of the true model size. Third, an active-set based algorithm is introduced to implement the constrained Dantzig selector efficiently. An appealing feature of this algorithm is that its convergence can be checked easily.

The rest of the paper is organized as follows. In Section \ref{Sec2}, we introduce the constrained Dantzig selector. We present its compressed sensing and sampling properties in Section \ref{Sec3}. In Section \ref{Sec4}, we discuss the implementation of the method and present several simulation and real data examples. We provide some discussions of our results and some possible extensions of our method in Section \ref{Sec5}. All technical details are relegated to the Appendix.\\

\section{The Constrained Dantzig selector} \label{Sec2}

To simplify the technical presentation, we adopt the model setting in Cand\`{e}s and Tao (2007) and present the main ideas focusing on the linear regression model
\begin{equation} \label{001}
\by = \bX \bbeta + \bveps,
\end{equation}
where $\by = (y_1, \ldots, y_n)\t$ is an $n$-dimensional response vector, $\bX = (\bx_1, \ldots, \bx_p)$ is an $n \times p$ design matrix consisting of $p$ covariate vectors $\bx_j$'s, $\bbeta = (\beta_1, \ldots, \beta_p)\t$ is a $p$-dimensional regression coefficient vector, and $\bveps = (\varepsilon_1, \ldots, \varepsilon_n)\t \sim N(\bzero, \sigma^2 I_n)$ for some positive constant $\sigma$ is an $n$-dimensional error vector independent of $\bX$. The normality assumption is considered for simplicity, and all the results in the paper can be extended to the cases of bounded errors or light-tailed error distributions without much difficulty. See, for example, the technical analysis in Fan and Lv (2011) and Fan and Lv (2013).

In both problems of compressed sensing and sparse modeling, we are interested in recovering the support and nonzero components of the true regression coefficient vector $\bbeta_0 = (\beta_{0,1}, \ldots, \beta_{0,p})\t$, which we assume to be sparse with $s$ nonzero components, for the case when the dimensionality $p$ may greatly exceed the sample size $n$. Throughout this paper, $p$ is implicitly understood as $\max\{n, p\}$ and $s \leq \min\{n, p\}$ to ensure model identifiability. To align the scale of all covariates, we assume that each column of $X$, that is, each covariate vector $\bx_j$, is rescaled to have $L_2$-norm $n^{1/2}$, matching that of the constant covariate vector $\textbf{1}$. The Dantzig selector (Cand\`{e}s and Tao, 2007) is defined as
\begin{equation} \label{002}
\hbbeta_{\text{DS}} = \argmin_{\bbeta \in \mathbb{R}^p} \ \|\bbeta\|_1 \ \text{ subject to } \ \|n^{-1}\bX\t (\by - \bX \bbeta)\|_\infty \leq \lambda_1,
\end{equation}
where $\lambda_1 \geq 0$ is a regularization parameter. The above constant Dantzig selector constraint on correlations between all covariates and the residual vector may not be flexible enough to differentiate important covariates and noise covariates. We introduce an extension of the Dantzig selector, the constrained Dantzig selector, defined as
\begin{equation} \label{003}
\begin{aligned}
\hbbeta_{\textsc{cds}} = & \argmin_{ {\bbeta \in \mathcal{B}_\lambda}} \ \|\bbeta\|_1\\
& \text{subject to } \ |n^{-1}\bx_j\t (\by - \bX \bbeta)| \leq \lambda_0 1_{\{|\beta_j| \geq \lambda\}} + \lambda_1 1_{\{|\beta_j|=0\}}\\
&
\text{for } j=1,\ldots,p,
\end{aligned}
\end{equation}
where $\lambda_0 \geq 0$ is a regularization parameter and $\mathcal{B}_\lambda = \{\bbeta \in \mathbb{R}^p: \beta_j = 0 \text{ or } |\beta_j| \geq \lambda \text{ for each } j\}$ is the constrained parameter space for some $\lambda \geq 0$. When we choose $\lambda = 0$ and $\lambda_0 = \lambda_1$, the constrained Dantzig selector becomes the Dantzig selector.
Throughout this paper, we choose the regularization parameters $\lambda_0$ and $\lambda_1$ as $c_0 \{(\log n) / n\}^{1/2}$ and $c_1 \{(\log p) / n\}^{1/2}$, respectively, with $\lambda_0 \leq \lambda_1$ as well as $c_0$ and $c_1$ two sufficiently large positive constants, and assume that $\lambda$ is a parameter greater than $\lambda_1$. The two parameters $\lambda_0$ and $\lambda_1$ differentially bound two types of correlations: on the support of the constrained Dantzig selector, the correlations between covariates and residuals are bounded, up to a common scale, by $\lambda_0$; on its complement, however, the correlations are bounded through $\lambda_1$.
In the ultra-high dimensional case, meaning $\log p = O(n^\alpha)$ for some {$0 < \alpha < 1$}, the constraints involving $\lambda_0$ are tighter than those involving $\lambda_1$, in which $\lambda_1$ is a universal regularization parameter for the Dantzig selector; see Cand\`{e}s and Tao (2007) and Bickel et al. (2009).

We now provide more insights into the new constraints in the constrained Dantzig selector. First, it is worthwhile to notice that if $\bbeta_0 \in \mathcal{B}_\lambda$, $\bbeta_0$ can satisfy new constraints with large probability in model setting (\ref{001}); see the proof of Theorem 1. With the tighter constraints, the feasible set of the constrained Dantzig selector problem is a subset of that of the Dantzig selector problem, resulting in a search of the solution in a reduced space. Second, it is appealing to extract more information in important covariates, leading to lower correlations between those variables and the residual vector. In this spirit, the constrained Dantzig selector puts tighter constraints on the correlations between selected variables and residuals. Third, the constrained Dantzig selector is defined on the constrained parameter space $\mathcal{B}_\lambda$, which has been introduced in Fan and Lv (2013). Such a space also shares some similarity to the union of coordinate subspaces considered in Fan and Lv (2011) for characterizing the restricted global optimality of nonconcave penalized likelihood estimators. The threshold $\lambda$ in $\mathcal{B}_\lambda$ distinguishes between important covariates with strong effects and noise covariates with weak effects. As shown in Fan and Lv (2013), this feature can lead to improved sparsity and effectively prevent overfitting by making it harder for noise covariates to enter the model.\\

\section{Main results} \label{Sec3}

In this section, two parallel theoretical results are introduced based on the uniform uncertainty principle (UUP) condition and restricted eigenvalue assumption, respectively. Although the UUP condition may be relatively stringent in some applications, we still present one theorem for our method under this condition in Section 3.1 for the purpose of comparison with the original Dantzig selector.

\subsection{Nonasymptotic compressed sensing properties} \label{Sec3.1}

Since the Dantzig selector was introduced partly for applications in compressed sensing, we first study the nonasymptotic compressed sensing properties of the constrained Dantzig selector by adopting the theoretical framework in Cand\`{e}s and Tao (2007). They introduced the uniform uncertainty principle condition defined as follows. Denote by $\bX_T$ a submatrix of $\bX$ consisting of columns with indices in a set $T \subset \{1, \ldots, p\}$. For the true model size $s$, define the $s$-restricted isometry constant of $\bX$ as the smallest constant $\delta_s$ such that
\[ 
(1-\delta_s) \|\bh\|_2^2 \leq n^{-1} \|\bX_T \bh\|_2^2 \leq (1+\delta_s) \|\bh\|_2^2
\] 
for any set $T$ with size at most $s$ and any vector $\bh$. This condition requires that each submatrix of $\bX$ with at most $s$ columns behaves similarly as an orthonormal system. Another constant, the $s$-restricted orthogonality constant, is defined as the smallest quantity $\theta_{s, 2s}$ such that
\[ 
n^{-1} |\langle \bX_T \bh, \bX_{T'}\bh' \rangle| \leq \theta_{s, 2s} \|\bh\|_2 \|\bh'\|_2
\] 
for all pairs of disjoint sets $T, T' \subset \{1, \ldots, p\}$ with $|T| \leq s$ and $|T'| \leq 2s$ and any vectors $\bh, \bh'$. The uniform uncertainty principle condition is simply stated as
\begin{equation} \label{L005}
\delta_s + \theta_{s, 2s} < 1.
\end{equation}
For notational simplicity, we drop the subscripts and denote these two constants by $\delta$ and $\theta$, respectively. Without loss of generality, assume that $\supp(\bbeta_0) = \{1 \leq j \leq p: \beta_{0, j} \neq 0\} = \{1, \ldots, s\}$ hereafter. To evaluate the sparse recovery accuracy, we consider the number of falsely discovered signs defined as $\FS(\hbbeta) = |\{j = 1, \ldots, p: \sgn(\hbeta_j) \neq \sgn(\beta_{0,j})\}|$ for an estimator $\hbbeta = (\hbeta_1, \ldots, \hbeta_p)\t$. Now we are ready to present the nonasymptotic compressed sensing properties of the constrained Dantzig selector.

\begin{theorem} \label{T1}
Assume that the uniform uncertainty principle condition (\ref{L005}) holds and $\bbeta_0 \in \mathcal{B}_\lambda$ with $\lambda \geq C^{1/2}(1+ \lambda_1/\lambda_0)\lambda_1$ for some positive constant $C$. Then with probability at least $1-O(n^{-c})$ for $c = (c_0^2 \wedge c_1^2)/(2\sigma^2) - 1$, the constrained Dantzig selector $\hbbeta$ satisfies that
{\begin{align}
\|\hbbeta - \bbeta_0\|_1 & \leq 2\sqrt{5}(1-\delta-\theta)^{-1} c_0 s \sqrt{(\log n) / n}, \nonumber \\
\|\hbbeta - \bbeta_0\|_2 & \leq \sqrt{5}(1-\delta-\theta)^{-1} c_0 \sqrt{s(\log n) / n}, \nonumber \\
\FS(\hbbeta) & \leq C c_1^2 s(\log p) / (n \lambda^2). \nonumber
\end{align}}
\hspace{-0.12in} If in addition {$\lambda > \sqrt{5}(1-\delta-\theta)^{-1} c_0 \sqrt{s(\log n) / n}$}, then with the same probability, it also holds that $\sgn(\hbbeta) = \sgn(\bbeta_0)$ and $\|\hbbeta - \bbeta_0\|_\infty \leq {2c_0 \|(n^{-1}\bX_1^T\bX_1)^{-1}\|_\infty \sqrt{(\log n) / n}}$, where $\bX_1$ is an $n \times s$ submatrix of $\bX$ corresponding to $s$ nonzero $\beta_{0,j}$'s.
\end{theorem}

The constant $c$ in the above probability bound can be sufficiently large since both constants $c_0$ and $c_1$ are assumed to be large, while the constant $C$ comes from Theorem 1.1 in Cand\`{e}s and Tao (2007); see the proof in the Appendix for details. In the above bound on the $L_\infty$-estimation loss, it holds that $\|(n^{-1} \bX_1\t \bX_1)^{-1}\|_\infty \leq s^{1/2} \|(n^{-1} \bX_1\t \bX_1)^{-1}\|_2 \leq (1 - \delta)^{-1} s^{1/2}$.
See Section 3.2 for more discussion on this quantity.

From Theorem 1, we see improvements of the constrained Dantzig selector over the Dantzig selector, which has a convergence rate, in terms of the $L_2$-estimation loss, up to a factor $\log p$ of that for the ideal procedure. However, the sparsity property of the Dantzig selector was not investigated in Cand\`{e}s and Tao (2007). In contrast, the constrained Dantzig selector is shown to have an inverse quadratic relationship between the number of falsely discovered signs and the threshold $\lambda$, revealing that its model selection accuracy increases with the signal strength. The number of falsely discovered signs can be controlled below or as an asymptotically vanishing fraction of the true model size, since $\FS(\hbbeta) \leq Cs(\lambda_1 / \lambda)^2 \leq (1+\lambda_1/\lambda)^{-2}s < s$ by assuming $\lambda > C^{1/2}(1+ \lambda_1/\lambda_0)\lambda_1$. 

Another advantage of the constrained Dantzig selector lies in its convergence rates. In the case of ultra-high dimensionality which is typical in compressed sensing applications, its prediction and estimation losses can be reduced from the logarithmic factor $\log p$ to $\log n$ with overwhelming probability. In particular, only a fairly weak assumption on the signal strength is imposed to attain such improved convergence rates. 
In fact, it has been shown in Raskutti et al. (2011) that without any  condition on the signal strength, the minimax convergence rate of $L_2$ risk has an upper bound of order $O\{s^{1/2}(\log p) /n\}$. Especially, they claimed that the Dantzig selector can achieve such minimax rate, but requires a relatively stronger condition on the design matrix than nonconvex optimization algorithms to determine the minimax upper bounds. We push a step forward that the constrained Dantzig selector, with additional signal strength conditions, can attain the $L_2$-estimation loss of a smaller order than $O\{s^{1/2}(\log p) /n\}$ in ultra-high dimensions.

There exist other methods which have been shown to enjoy convergence rates of the same order as well, for example, in Zheng et al. (2014) for high-dimensional thresholded regression. However, these results usually rely on a stronger condition on signal strength, such as, the minimum signal strength is at {least} of order $\{s(\log p)/n\}^{1/2}$. In another work, Fan and Lv (2011) showed that the nonconcave penalized estimator can have a consistency rate of $O_p(s^{1/2}n^{-\gamma} \log n)$ for some $\gamma \in (0, 1/2]$ under the $L_2$-estimation loss, which can be slower than our rate of convergence. More detailed discussion on the relationship between the faster rates of convergence and the assumptions on the signal strength and sparsity can be found in Section \ref{Sec3.2}. A main implication of our improved convergence rates is that a smaller number of observations will be needed for the constrained Dantzig selector to attain the same level of accuracy in compressed sensing, as the Dantzig selector, which is demonstrated in Section \ref{Sec4.2}.

\subsection{Sampling properties} \label{Sec3.2}

The properties of the Dantzig selector have also been extensively investigated in Bickel et al. (2009). They introduced the restricted eigenvalue assumption with which the oracle inequalities under various prediction and estimation losses were derived. We adopt their theoretical framework and study the sampling properties of the constrained Dantzig selector under the restricted eigenvalue assumption stated below. {A positive integer $m$ is said to be in the same order of $s$ if $m/s$ can be bounded from both above and below by some positive constants}.

\begin{assumption} \label{C1}
For some positive integer $m$ in the {same} order of $s$, there exists some positive constant $\kappa$ such that $\|n^{-1/2}\bX \bdelta\|_2 \geq \kappa \max\{\|\bdelta_{1}\|_2, \|\bdelta_1'\|_2\}$ for all $\bdelta \in \mathrm{R}^p$ satisfying $\|\bdelta_{2}\|_1 \leq \|\bdelta_{1}\|_1$, where $\bdelta = (\bdelta_1\t, \bdelta_2\t)\t$, $\bdelta_1$ is a subvector of $\bdelta$ consisting of the first $s$ components, and $\bdelta_1'$ is a subvector of $\bdelta_2$ consisting of the $\max\{m, C_m s(\lambda_1 / \lambda)^2\}$ largest components in magnitude, with $C_m$ some positive constant.
\end{assumption}

Condition \ref{C1} is a basic assumption on the design matrix $\bX$ for deriving the oracle inequalities of the Dantzig selector. {Since we assume that $\supp(\bbeta_0) = \{1, \ldots, s\}$, Condition \ref{C1} indeed plays the same role as the restricted eigenvalue assumption $\mathrm{RE}(s,m,1)$ in Bickel et al. (2009), which assumes the inequality in Condition \ref{C1} holds for any subset with size no larger than $s$ to cover all possibilities of $\supp(\bbeta_0)$.} See (\ref{A004}) in the Appendix for insights into the basic inequality $\|\bdelta_{2}\|_1 \leq \|\bdelta_{1}\|_1$ and Bickel et al. (2009) for more detailed discussions on this assumption.

\begin{theorem} \label{T2}
Assume that Condition \ref{C1} holds and $\bbeta_0 \in \mathcal{B}_\lambda$ with $\lambda \geq C_m^{1/2}(1+ \lambda_1/\lambda_0)\lambda_1$. Then the constrained Dantzig selector $\hbbeta$ satisfies with the same probability as in Theorem \ref{T1} that
{\begin{align}
n^{-1/2}\|\bX(\hbbeta - \bbeta_0)\|_2 & = O(\kappa^{-1} \sqrt{s (\log n) / n}), \quad \|\hbbeta - \bbeta_0\|_1 = O(\kappa^{-2}s \sqrt{(\log n) / n}), \nonumber \\
\|\hbbeta - \bbeta_0\|_2 & = O(\kappa^{-2} \sqrt{s(\log n) / n}), \quad \FS(\hbbeta) \leq C_m c_1^2 s(\log p) / (n \lambda^2). \nonumber
\end{align}}
\hspace{-0.12in} If in addition {$\lambda > 2\sqrt{5} \kappa^{-2}c_0 s^{1/2}\sqrt{(\log n) / n}$}, then with the same probability it also holds that $\sgn(\hbbeta) = \sgn(\bbeta_0)$ and {$\|\hbbeta - \bbeta_0\|_\infty = O\{ \|(n^{-1}\bX_1^T \bX_1)^{-1}\|_\infty \sqrt{(\log n) / n}\}$.}
\end{theorem}

Theorem \ref{T2} establishes asymptotic results on the sparsity and oracle inequalities for the constrained Dantzig selector under the restricted eigenvalue assumption. This assumption, which is an alternative to the uniform uncertainty principle condition, has also been widely employed in high-dimensional settings. In Bickel et al. (2009), an approximate equivalence of the Lasso estimator (Tibshirani, 1996) and Dantzig selector was proved under this assumption, and the Lasso estimator was shown to be sparse with size $O(\phi_{\max} s)$, where $\phi_{\max}$ is the largest eigenvalue of the Gram matrix $n^{-1} \bX\t \bX$. In contrast, the constrained Dantzig selector gives a sparser model under the restricted eigenvalue assumption, since its number of falsely discovered signs $\FS(\hbbeta) \leq C_m s(\lambda_1 / \lambda)^2 = o(s)$ when $\lambda_1 = o(\lambda)$. Similar as in Theorem 1, the constrained Dantzig selector improves over both Lasso and the Dantzig selector in terms of convergence rates, a reduction of the $\log p$ factor to $\log n$.

Let us now take a closer look at the relationship between the faster rates of convergence and the assumptions on the signal strength and sparsity. Observe that the enhanced consistency results require the minimum signal strength to be at least of order $(\log p) / \sqrt{n \log n}$, in view of the assumption $\lambda \geq C_m^{1/2}(1+ \lambda_1/\lambda_0)\lambda_1$. 
Assume for simplicity that $\|\bbeta_0\|_2$ is bounded away from both zero and $\infty$. Then the order of the minimum signal strength yields an upper bound on the sparsity level $s = O\{n (\log n) / (\log p)^2\}$, which means that the sparsity $s$ is required to be smaller when the dimensionality $p$ becomes larger. In the ultra-high dimensional setting of $\log p = O(n^\alpha)$ with $0 < \alpha < 1$, we then have $s = O(n^{1 - 2\alpha} \log n)$. Thus the classical convergence rates involving $s (\log p) /n = O(n^{-\alpha} \log n)$ still go to zero asymptotically, while the rates established in our paper are improved with $s (\log p) /n$ replaced by $s (\log n) /n = O\{n^{-2\alpha} (\log n)^2\}$. We see a gain on the convergence rate of a factor $n^{\alpha}/(\log n)$ at the cost of the aforementioned signal strength and sparsity assumptions.

One can see that results in Theorems 1 and 2 are approximately equivalent, while the latter presents an additional oracle inequality on the prediction loss. An interesting phenomenon is that if adopting a simpler version, that is, the Dantzig selector equipped with the thresholding constraint only, we can also obtain similar results as in Theorems 1 and 2, but with a stronger condition on signal strength such as $\lambda \gg s^{1/2} \lambda_1$. In this sense, the constrained Dantzig selector is also an extension of the Dantzig selector equipped with the thresholding constraint only, but enjoys better properties. Some comprehensive results on the prediction and variable selection properties have also been established in Fan and Lv (2013) for various regularization methods, revealing their asymptotic equivalence in the thresholded parameter space. However, as mentioned in Section 3.1, improved rates as in Theorem 2 commonly require a stronger assumption on the signal strength, which is $\bbeta_0 \in \mathcal{B}_\lambda$ with $\lambda \gg s^{1/2}\lambda_1$; see, for example, Theorem 2 of Fan and Lv (2013).

For the quantity $\|(n^{-1}\bX_1^T \bX_1)^{-1}\|_\infty$ in the above bound on the $L_\infty$-estimation loss, if $\bX_1$ takes the form of a common correlation matrix $(1 - \rho) I_s + \rho \textbf{1}_s \textbf{1}_s\t$ for some $\rho \in [0, 1)$, it is easy to check that $\|(n^{-1}\bX_1^T \bX_1)^{-1}\|_\infty = (1 - \rho)^{-1} \{1 + (2 s - 3) \rho\}/\{1 + (s - 1) \rho\}$, which is bounded regardless of the value of $s$.

\subsection{Asymptotic properties of computable solutions} \label{Sec3.3}

The nonasymptotic and sampling properties of the constrained Dantzig selector, as the global minimizer, have been established in Sections 3.1 and 3.2, respectively. However, it is not guaranteed that the global minimizer can be generated by a computational algorithm. Moreover, a computable solution, generated by any algorithm, may only be a local minimizer in many cases. Under certain regularity conditions, we demonstrate that the local minimizer of our method can still share the same nice asymptotic properties as the global one.


\begin{theorem} \label{T3}

Let $\hbbeta$ be a computable local minimizer of (\ref{003}). 
Assume that there exist some positive constants $c_2$, $\kappa_0$ and sufficiently large positive constant $c_3$ such that $\|\hbbeta\|_0 \leq c_2 s$ and $\min_{\|\bdelta\|_2 = 1,\ \|\bdelta\|_0 \leq c_3 s} n^{-1/2} \|\bX \bdelta\|_2 \geq \kappa_0$. Then under conditions of Theorem \ref{T2}, $\hbbeta$ enjoys the same properties as the global minimizer in Theorem \ref{T2}.

\end{theorem}


In Section 4.1, we introduce an efficient algorithm that gives us a local minimizer. Theorem 3 indicates that the obtained solution can also enjoy the asymptotic properties in Theorem \ref{T2} under the extra assumptions that $\hbbeta$ is a sparse solution with the number of nonzero components comparable with $s$ and the design matrix $\bX$ satisfies a sparse eigenvalue condition. 
Similar results for the computable solution can be found in Fan and Lv (2013, 2014), where the local minimizer is additionally assumed to satisfy certain constraint on the correlation between the residual vector and all the covariates.\\


\section{Numerical studies} \label{Sec4}

In this section, we first introduce an algorithm which can efficiently implement the constrained Dantzig selector. Then several simulation studies and two real data examples are presented to evaluate the performance of our method.

\subsection{Implementation} \label{Sec4.1}

The constrained Dantzig selector defined in (\ref{003}) depends on tuning parameters $\lambda_0$, $\lambda_1$, and $\lambda$. {We suggest some fixed values for $\lambda_0$ and $\lambda$ to simplify the computation, {since the proposed method is generally not that sensitive to $\lambda_0$ and $\lambda$ as long as they fall in certain ranges. In simulation studies to be presented,} a value around $\{(\log p)/n\}^{1/2}$ for $\lambda$ and {a smaller value for $\lambda_0$, say $0.05\{(\log p)/n\}^{1/2}$ or $0.1\{(\log p)/n\}^{1/2}$, can provide us nice prediction and estimation results.
The value of $\lambda_0$ is chosen to be smaller than $\{(\log p)/n\}^{1/2}$ to mitigate the selection of noise variables and facilitate sparse modeling. The performance of our method with respect to different values of $\lambda_0$ and $\lambda$ is shown in simulation Example 2 in Section 4.2, which is a typical example that illustrates the robustness of the proposed method with respect to $\lambda_0$ and $\lambda$.}}

For fixed $\lambda_0$ and $\lambda$, we exploit the idea of sequential linear programming to produce the solution path of the constrained Dantzig selector as $\lambda_1$ varies. Choose a grid of values for the tuning parameter $\lambda_1$ in decreasing order with the first one being $\|n^{-1}\bX^T \by\|_\infty$. It is easy to check that $\bbeta = \bzero$ satisfies all the constraints in (\ref{003}) for $\lambda_1 = \|n^{-1}\bX\t \by\|_\infty$, and thus the solution is $\hbbeta_{\textsc{cds}} = \bzero$ in this case. For each $\lambda_1$ in the grid, we use the solution from the previous one in the grid as an initial value to speed up the convergence. For a given $\lambda_1$, we define an active set, iteratively update this set, and solve the constrained Dantzig selector problem. We name this algorithm as the CDS algorithm which is detailed in four steps below.
\begin{enumerate}
  \item For a fixed $\lambda_1$ in the grid, denote by $\hbbeta_{\lambda_1}^{(0)}$ the initial value. Let $\hbbeta_{\lambda_1}^{(0)}$ be zero when $\lambda_1 = \|n^{-1}\bX\t \by\|_\infty$, and the estimate from previous $\lambda_1$ in the grid otherwise.
  \item Denote by $\hbbeta_{\lambda_1}^{(k)}$ the estimate from the $k$th iteration. Define the active set $\mathcal{A}$ as the support of $\hbbeta_{\lambda_1}^{(k)}$ and $\mathcal{A}^c$ its complement. Let $\bb$ be a vector with constant components $\lambda_0$ on $\mathcal{A}$ and $\lambda_1$ on $\mathcal{A}^c$. For the $(k+1)$th iteration, update $\mathcal{A}$ as $\mathcal{A} \cup \{j \in \mathcal{A}^c : |n^{-1} \bx\t_j (\by - \bX \hbbeta_{\mathcal{A}})| > \lambda_1\}$, where the subscript $\mathcal{A}$ indicates a subvector restricted on $\mathcal{A}$. Solve the following linear program on the new set $\mathcal{A}$:
\begin{equation} \label{007}
        \hbbeta_{\mathcal{A}} = \argmin \ \|\bbeta_{\mathcal{A}}\|_1 \text{ subject to } \ |n^{-1}\bX\t_{\mathcal{A}} (\by - \bX_{\mathcal{A}}
        \bbeta_{\mathcal{A}})| \preceq \bb_{\mathcal{A}},
\end{equation}
where $\preceq$ is understood as componentwise no larger than and the subscript $\mathcal{A}$ also indicates a submatrix with columns corresponding to $\mathcal{A}$. For the solution obtained in (\ref{007}), set all its components smaller than $\lambda$ in magnitude to zero.
  \item Update the active set $\mathcal{A}$ as the support of $\hbbeta_{\mathcal{A}}$. Solve the Dantzig selector problem on this active set with $\lambda_0$ as the regularization parameter:
\begin{equation} \label{008}
        \hbbeta_{\mathcal{A}} = \argmin \ \|\bbeta_{\mathcal{A}}\|_1 \text{ subject to } \|n^{-1}\bX\t_{\mathcal{A}} (\by - \bX_{\mathcal{A}} \bbeta_{\mathcal{A}})\|_\infty \leq \lambda_0.
\end{equation}
Let $\hbbeta_{\mathcal{A}}^{(k+1)} =\hbbeta_{\mathcal{A}}$ and $\hbbeta_{\mathcal{A}^c}^{(k+1)} = \bzero$, which give the solution for the $(k+1)$th iteration.
  \item Repeat steps 2 and 3 until convergence for a fixed $\lambda_1$ and record the estimate from the last iteration as $\hbbeta_{\lambda_1}$. Jump to the next $\lambda_1$ if $\hbbeta_{\lambda_1} \in \mathcal{B}_\lambda$, and stop the algorithm otherwise.
\end{enumerate}

With the solution path produced, we use the cross-validation to select the tuning parameter $\lambda_1$. {One can also tune $\lambda_0$ and $\lambda$ similarly as for $\lambda_1$, but as suggested before, some fixed values for them suffice to obtain satisfactory results.}

The rationales of the constrained Dantzig selector algorithm are as follows. Step 1 defines the initial value ($0$th iteration) for each $\lambda_1$ in the grid. In step 2, starting with a smaller active set, we add variables that violate the constrained Dantzig selector constraints to eliminate such conflict. As a consequence, some components of $\bb_{\mathcal{A}}$ are of value $\lambda_1$ instead of $\lambda_0$. Therefore, we need to further solve (\ref{008}) in step 3 by noting that restricted on its support, the constrained Dantzig selector should be a solution to the Dantzig selector problem with parameter $\lambda_0$. An early stopping of the solution path is imposed in step 4 to make this algorithm computationally more efficient.

An appealing feature of this algorithm is that its convergence can be checked easily. Once there are no more variables violating the constrained Dantzig selector constraints, that is, $\{j \in \mathcal{A}^c : |n^{-1} \bx\t_j (\by - \bX \hbbeta_{\mathcal{A}})| > \lambda_1\} = \emptyset$, the iteration stops and the algorithm converges. In other words, the convergence of the algorithm is equivalent to that of the active set which can be checked directly. When the algorithm converges, the solution lies in the feasible set of the constrained Dantzig selector problem and is a global minimizer restricted on the active set, and is thus a local minimizer.

In simulation {Example} 2 of Section 4.2, we tracked the convergence property of the algorithm on $100$ data sets for $p=1000$, $5000$, and {$10000$}, respectively. In all cases, we observe that the algorithm always converged over all $100$ simulations, indicating considerable stability of this algorithm. Another advantage of the algorithm is that it is built upon the Dantzig selector in lower dimensions, so it inherits the computational efficiency.\\

\subsection{Simulation studies} \label{Sec4.2}

To better illustrate the performance of the constrained Dantzig selector, we consider the thresholded Dantzig selector which simply sets components of the Dantzig selector estimate to zeros if smaller than a threshold in magnitude. We evaluated the performance of the constrained Dantzig selector in comparison with the Dantzig selector, thresholded Dantzig selector, Lasso, elastic net (Zou and Hastie, 2005), and adaptive Lasso (Zou, 2006). Two simulation studies were considered, with the first one investigating sparse recovery for compressed sensing and the second one examining sparse modeling.

The setting of the first simulation {example} is similar to that of the sparse recovery example in Lv and Fan (2009). The noiseless sparse recovery example is considered here since the Dantzig selector problem originated from compressed sensing. We want to evaluate the capability of our constrained Dantzig selector in recovering sparse signals as well. We generated $100$ data sets from model (\ref{001}) without noise, that is, the linear equation $\by = \bX \bbeta_0$ with $(s, p)=(7, 1000)$. The nonzero components of $\bbeta_0$ were set to be $(1,\ -0.5,\ 0.7,\ -1.2,\ -0.9,\ 0.3,\ 0.55)\t$ lying in the first seven components, and $n$ was chosen to be even integers between $30$ and $80$. The rows of the design matrix $\bX$ were sampled as independent and identically distributed (i.i.d.) copies from $N(\bzero, \bGamma_r)$, with $\bGamma_r$ a $p \times p$ matrix with diagonal elements being $1$ and off-diagonal elements being $r$, and then each column was rescaled to have $L_2$-norm {$\sqrt{n}$}. Three levels of population collinearity, $r=0$, $0.2$, and $0.5$, were considered. Let $\lambda_0$ and $\lambda$ be in two small grids $\{0.001$, $0.005$, $0.01$, $0.05$, $0.1\}$ and $\{0.05$, $0.1$, $0.15$, $0.2\}$, respectively. {We chose two grids of values for them since in the literature of
compressed sensing, it is desirable to have the true support included among a set of estimates. The value $0.2$ was chosen because it is close to $\{(\log p)/n\}^{1/2}$, which is about $\{(\log 1000)/80\}^{1/2}$ in this example. Smaller values for $\lambda$ and $\lambda_0$ are also included in the grid for conservativeness.} We set the grid of values for $\lambda_1$ as described in Section 4.1. If any of the solutions in the path had exactly the same support as $\bbeta_0$, it is counted as successful recovery. This criterion was applied to all other methods in this example for fair comparison.

\begin{figure}[!htbp] \centering
\begin{center}%
\begin{tabular}
[l]{l}%
{\hspace{-0.1in}\includegraphics[scale=0.55]{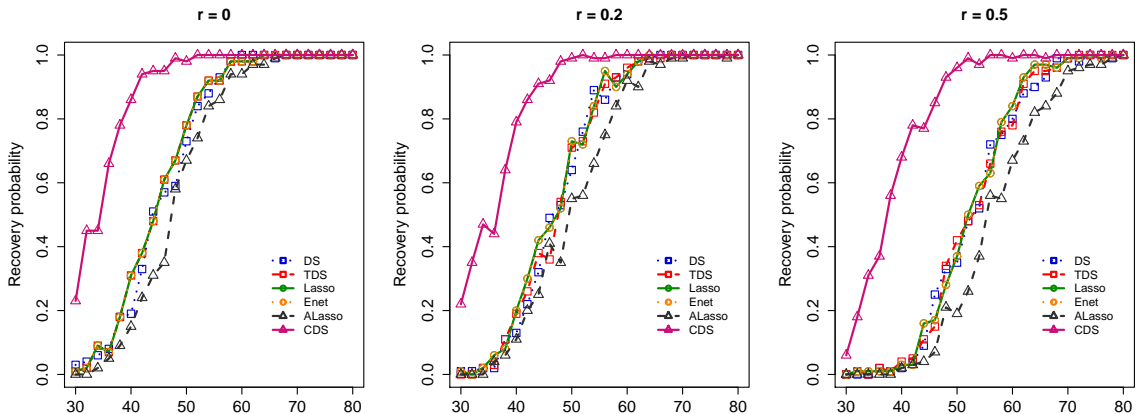}
}
\end{tabular}
\vspace{-0.2in}
\caption{Probabilities of exact sparse recovery for the Dantzig selector (DS), thresholded Dantzig selector (TDS), Lasso, elastic net (Enet), adaptive Lasso (ALasso), and constrained Dantzig selector (CDS) in simulation {example} 1 of Section 4.2.}
\label{Fig1}%
\end{center}%
\end{figure}

Figure \ref{Fig1} presents the probabilities of exact recovery of sparse $\bbeta_0$ based on $100$ simulations by all methods. We see that all methods performed well in relatively large samples and had lower probability of successful sparse recovery when the sample size becomes smaller. The constrained Dantzig selector performed better than other methods over different sample sizes and three levels of population collinearity. In particular, the thresholded Dantzig selector performed similarly to the Dantzig selector, revealing that simple thresholding alone, instead of flexible constraints as in the constrained Dantzig selector, does not help much on signal recovery in this case.

The second simulation {example} adopts a similar setting to that in Zheng et al. (2014). We generated $100$ data sets from the linear regression model (\ref{001}) with Gaussian error $\bveps \sim N(\bzero,\sigma^2 I_n)$. The coefficient vector $\bbeta_0 = (\bv^T, \ldots, \bv^T, \textbf{0}\t)\t$ with the pattern $\bv = (\bbeta_{\text{strong}}\t, \bbeta_{\text{weak}}\t)\t$ repeated three times, where $\bbeta_{\text{strong}} = (0.6, 0, 0, -0.6, 0, 0)\t$ and  $\bbeta_{\text{weak}} = (0.05, 0, 0, -0.05, 0, 0)\t$. The coefficient subvectors $\bbeta_{\text{strong}}$ and $\bbeta_{\text{weak}}$ stand for the strong signals and weak signals in $\bbeta_0$, respectively. The sample size and noise level were chosen as $(n, \sigma) = (100, 0.4)$, while the {dimensionality} $p$ was set to be {$1000$, $5000$, and $10000$}. The rows of the $n \times p$ design matrix $\bX$ were sampled as i.i.d.
copies from a multivariate normal distribution $N(\bzero, \bSig)$ with $\bSig = (0.5^{|i-j|})_{1 \leq i, j \leq p}$. {We applied all methods as in simulation example 1 and set $\lambda_0 = 0.01$ and $\lambda = 0.2$ for our method. Similarly as before, the value of $0.2$ was selected since it is close to $\{(\log 1000)/100\}^{1/2}$}. The ideal procedure, which knows the true underlying sparse model in advance, was also used as a benchmark.

\begin{table}[!htb]
\caption{Means and standard errors (in parentheses) of different performance measures by all methods in simulation {example} 2 of Section 4.2.}
\centering
\tabcolsep=0.11cm
\scalebox{0.9}{
\begin{tabular}{llcccccc}
\hline
Measure &      DS  &   TDS   &    Lasso &   Enet  &  ALasso  & CDS &  Oracle  \smallskip \\
\hline
$p = 1000$ & &  &  &  &  &  &  \\
PE ($\times 10^{-2}$) & 30.8 (0.6) & 28.5 (0.5) & 30.3 (0.5) & 32.9 (0.7) & 19.1 (0.1) & 18.5 (0.1) & 18.2 (0.1) \\
$L_1$ ($\times 10^{-2}$) & 201.9 (5.4) & 137.2 (3.4) & 186.1 (5.0) & 211.1 (6.0) & 58.0 (1.1) & 51.3 (0.6) & 41.5 (0.9) \\
$L_2$ ($\times 10^{-2}$) & 40.1 (0.7) & 37.2 (0.7) & 39.8 (0.7) & 43.0 (0.8) & 18.3 (0.3) & 16.3 (0.2) & 14.7 (0.3) \\
$L_{\infty}$ ($\times 10^{-2}$) & 19.0 (0.5) & 18.6 (0.4) & 19.3 (0.5) & 20.7 (0.5) & 9.1 (0.3) & 7.5 (0.2) & 8.4 (0.2) \\
FP & 44.4 (1.8) & 5.5 (3.8) & 36.3 (1.6) & 44.1 (1.9) & 0.5 (0.1) & 0 (0) & 0 (0) \\
FN.strong & 0 (0) & 0 (0) & 0 (0) & 0 (0) & 0 (0) & 0 (0) & 0 (0) \\
FN.weak & 5.3 (0.1) & 5.9 (0.4) & 5.4 (0.1) & 5.3 (0.1) & 6.0 (0.0) & 6.0 (0.0) & 0 (0)
    \smallskip \\
    \hline
$p = 5000$ & &  &  &  &  &  &  \\
PE ($\times 10^{-2}$) & 45.1 (1.1) & 39.3 (1.1) & 44.8 (1.1) & 44.9 (1.1) & 21.3 (0.6) & 18.4 (0.1) & 18.3 (0.1) \\
$L_1$ ($\times 10^{-2}$) & 289.3 (6.4) & 184.6 (4.5) & 270.8 (6.8) & 273.2 (6.9) & 71.2 (2.1) & 50.4 (0.7) & 41.7 (1.1) \\
$L_2$ ($\times 10^{-2}$) & 56.3 (1.1) & 50.6 (1.1) & 56.1 (1.1) & 56.2 (1.1) & 22.9 (0.9) & 16.0 (0.2) & 14.9 (0.4) \\
$L_{\infty}$ ($\times 10^{-2}$) & 27.4 (0.7) & 25.1 (0.6) & 27.8 (0.7) & 27.8 (0.7) & 12.7 (0.7) & 7.3 (0.2) & 8.8 (0.3) \\
FP & 60.6 (1.7) & 7.0 (3.5) & 53.5 (2.3) & 53.8 (2.1) & 1.0 (0.1) & 0 (0) & 0 (0) \\
FN.strong & 0 (0) & 0 (0) & 0 (0) & 0 (0) & 0 (0) & 0 (0) & 0 (0) \\
FN.weak & 5.2 (0.1) & 6.0 (0.1) & 5.4 (0.1) & 5.3 (0.1) & 5.9 (0.0) & 6.0 (0.0) & 0 (0)\\
    \hline
    {$p = 10000$} & &  &  &  &  &  &  \\
PE ($\times 10^{-2}$) & 54.4 (16.2) & 54.8 (17.4) & 56.7 (23.1) & 63.3 (21.7) & 32.4 (32.3) & 19.0 (0.5) & 18.3 (0.1) \\
$L_1$ ($\times 10^{-2}$) & 322.3 (67.9) & 218 (60.3) & 296.3 (72.7) & 334.8 (79.9) & 86.6 (67.6) & 51.9 (1.6) & 41.3 (0.9) \\
$L_2$ ($\times 10^{-2}$) & 62.7 (14.1) & 63.5 (14.9) & 64.2 (17.7) & 70.3 (16.8) & 29.1 (26.3) & 16.6 (0.6) & 14.7 (0.3) \\
$L_{\infty}$ ($\times 10^{-2}$) & 30.5 (8.2) & 32.5 (8.5) & 31.8 (9.5) & 34.4 (9.3) & 15.8 (13.2) & 7.8 (0.6) & 8.5 (0.2) \\
FP & 69.9 (14.3) & 6.9 (4.6) & 56.89 (20.66) & 64.06 (18.64) & 0.78 (1.25) & 0.01 (0.10) & 0 (0) \\
FN.strong & 0 (0.1) & 0 (0.1) & 0.02 (0.14) & 0.02 (0.14) & 0.21 (0.95) & 0.01 (0.10) & 0 (0) \\
FN.weak & 6 (0.2) & 6 (0.1) & 5.98 (0.14) & 5.97 (0.17) & 6 (0) & 6 (0) & 0 (0)\\
    \hline
\end{tabular}
}
\label{Tab1}
\end{table}

To compare these methods, we considered several performance measures: the prediction error, the $L_q$-estimation loss with $q = 1, 2, \infty$, number of false positives, and number of false negatives for strong or weak signals. The prediction error is defined as $E (Y - \bx\t \hbbeta)^2$ with $\hbbeta$ an estimate and $(\bx\t, Y)$ an independent observation, and the expectation was calculated using an independent test sample of size $10000$. A false positive means a falsely selected noise covariate in the model, while a false negative means a missed true covariate.
Table \ref{Tab1} summarizes the comparison results by all methods. We observe that most weak covariates were missed by all methods. This is reasonable since the weak signals are around the noise level, making it difficult to distinguish them from the noise covariates. However, the constrained Dantzig selector outperformed other methods in terms of other prediction and estimation measures, and followed very closely the ideal procedure in all cases of $p=1000$, {$5000$, and $10000$}. In particular, the $L_\infty$-estimation loss for the constrained Dantzig selector was similar to that for the oracle procedure, confirming tight bounds on this loss in Theorems 1 and 2. As the {dimensionality} grows {higher}, the constrained Dantzig selector performed similarly, while other methods suffered from high dimensionality. In particular, the thresholded Dantzig selector has been shown to improve over the Dantzig selector, but was still outperformed by the adaptive Lasso and constrained Dantzig selector in this {example}, revealing the necessity to introduce more flexible constraints instead of simple thresholding.

{Recall that we have fixed $\lambda_0=0.01$ and $\lambda=0.2$ in the simulation example 2 across all settings. {We now study the robustness of the constrained Dantzig with respect to $\lambda_0$ and $\lambda$ in this typical example. For simplicity, we only consider Example 2 with dimensionality $p=1000$}. Instead of fixing $\lambda_0=0.01$ and $\lambda=0.2$, we let $\lambda_0$ be a value in {the grid} $\{0.001, 0.005, 0.01, 0.015, 0.02, 0.025, 0.03\}$ and $\lambda$ a value {in} $\{0.1, 0.15, 0.2, 0.25, 0.3, 0.35\}$, respectively. Therefore, we study $7 \times 6 = 42$ different combinations of $\lambda_0$ and $\lambda$ in total to evaluate the robustness. The same performance measures 
were calculated. Here we only present the prediction error results to save space, and the other results are available upon request. The {tuning} parameter $\lambda_1$ was chosen by cross-validation, similar as in Example 2. We can see from Table 2 that for all {$\lambda \geq 0.15$} and all $\lambda_0$ in the grids, the means and standard errors of the prediction error {are} very close or {even} identical to those for $\lambda_0=0.01$ and $\lambda=0.2$. {The prediction error in the case of $\lambda = 0.1$ is slightly higher since when the threshold becomes lower, some noise variables  can be included.}
For the results on estimation {losses} as well as false positives and false negatives, we observe similar patterns confirming the robustness of our method {with respect} to the {choices} of $\lambda_0$ and $\lambda$.}

\begin{table}[!htb] \label{robust}
\caption{\small{Means and standard errors (in parentheses) of prediction error of the constrained Dantzig selector for different choices of $\lambda_0$ and $\lambda$ in simulation example 2 {with} $p=1000$.}}
\centering
\tabcolsep=0.15cm
\scalebox{0.9}{
\begin{tabular}{ l ccccccc }
\hline
	& \multicolumn{6}{c}{$\lambda$} \\
	\cmidrule(lr){2-7}
	$\lambda_0$ & 0.10 & 0.15 & 0.20 & 0.25 & 0.30 & 0.35 \\ 
	\cmidrule(lr){2-7}
	0.001 & 0.192 (0.002) & 0.186 (0.001) & 0.185 (0.001) & 0.185 (0.001) & 0.185 (0.001) & 0.185 (0.001) \\
	0.005 & 0.191 (0.003) & 0.188 (0.001) & 0.185 (0.001) & 0.185 (0.001) & 0.185 (0.001) & 0.185 (0.001) \\
	0.010 & 0.192 (0.002) & 0.187 (0.001) & 0.185 (0.001) & 0.185 (0.001) & 0.188 (0.003) & 0.185 (0.001) \\
	0.015 & 0.196 (0.007) & 0.186 (0.001) & 0.185 (0.001) & 0.185 (0.001) & 0.185 (0.001) & 0.185 (0.001) \\
	0.020 & 0.192 (0.002) & 0.188 (0.001) & 0.185 (0.001) & 0.185 (0.001) & 0.185 (0.001) & 0.185 (0.001) \\
	0.025 & 0.190 (0.002) & 0.187 (0.001) & 0.185 (0.001) & 0.185 (0.001) & 0.185 (0.001) & 0.185 (0.001) \\
	0.030 & 0.191 (0.002) & 0.187 (0.001) & 0.191 (0.007) & 0.188 (0.003) & 0.185 (0.001) & 0.185 (0.001) \\
\hline
\end{tabular}
}
\end{table}

\subsection{Real data analyses} \label{Sec4.3}

{
We applied the same methods as in Section 4.2 to two real data sets: one real PCR data {set} and another gene expression data {set,}
both {in the high-dimensional setting with relatively small sample size}. In both data sets, we found that the proposed method {enjoys smaller} prediction errors and the differences {are} statistically significant.

The real PCR data set, {originally studied in Lan et al. (2006), examines} the genetics of two inbred mouse populations. This data set is comprised of $n = 60$ samples with $29$ {males} and $31$ females. Expression levels of $22,575$ genes were measured. Following Song and Liang (2015), we study the linear relationship between the numbers of Phosphoenolpyruvate carboxykinase (PEPCK), a phenotype measured by quantitative real-time PCR, and the gene expression levels. Both the phenotype and predictors are continuous. As suggested in Song and Liang (2015), we only picked $p = 2000$ genes having the highest marginal {correlations} with PEPCK as predictors. 
The response was standardized to have zero mean and unit variance before we conducted the analysis. The $2000$ predictors were standardized to have zero mean and $L_2$-norm $\sqrt{n}$ in each column. 
Then the data set was randomly split into a training {set} of $55$ samples and {a test set with the remaining $5$ samples} for $100$ times. We set $\lambda_0=0.001$ and $\lambda=0.02$ in this real data {analysis as well as the other one below for conservativeness.}
Methods under comparison are the same as in the {simulation} studies.

Table 3 reports the means and standard errors of the prediction error on the test data as well as the median model size for each method. 
We see that the method CDS gave the lowest mean prediction error.
Paired $t$-tests were conducted {for} the prediction errors {of} CDS versus DS, TDS, Lasso, ALasso, and Enet, respectively, to {test the differences in performance across various methods}. The corresponding $p$-values were 0.0243, 0.0014, 0.0081, 0.0001, and 0.0102, respectively, indicating {significantly} different prediction {error from that for} CDS.

\begin{table}[!htb] \label{gse3330}
\caption{Means and standard errors of the prediction error and median model size over $100$ random splits for the real PCR data set.}
\centering
\begin{tabular}{ ccc }
\hline
	Method & PE & Model Size \\ \hline
	DS & 0.773 (0.040) & 54 \\
	TDS & 0.897 (0.063) & 14 \\
	Lasso & 0.802 (0.043) & 58 \\
	ALasso & 0.922 (0.056) & 5 \\
	Enet & 0.793 (0.041) & 58 \\
	CDS & 0.660 (0.036) & 21 \\ \hline
\end{tabular}
\end{table}


The second data set has been studied in Scheetz et al. (2006) and Huang et al. (2008). In this data set, $120$ twelve-week-old male rats were selected for tissue harvesting from the eyes and for microarray analysis. There are $31,042$ different probe sets in the microarrays from the RNA of those rat eyes. Following Huang et al. (2008), we excluded the probes that were not expressed sufficiently or that lacked sufficient variation, leaving $18,976$ probes which satisfy these two criteria. The response variable TRIM32, which was recently found to cause Bardet-Biedl syndrome, is one of the selected $18,976$ probes. We then selected 3,000 probes with the largest variances from the remaining $18,975$ probes. The goal of our analysis is to identify the genes that are most relevant to the expression level of TRIM32 from the $3,000$ candidate genes.

{Similarly as in the analysis of} the first real data set, {we standardized the response and predictors beforehand.}
The training set contains $100$ samples and was sampled randomly $100$ times from the full data set. The remaining $20$ samples at each time served as the test set. {Results of prediction errors and median model sizes are presented in Table 4}. {It is clear that the proposed method CDS enjoys the lowest prediction error with a small model size.} We conducted the same paired $t$-tests as in the real PCR data set for comparison. The corresponding $p$-values were 0.0351, 0.0058, 0.0164, 0.0004, and 0.0344, respectively, showing significant improvement. 

\begin{table}[!htb] \label{rateyes}
\caption{Means and standard errors of the prediction error and median model size over $100$ random splits for the gene expression data set of rat eyes.}
\centering
\begin{tabular}{ ccc }
\hline
	Method & PE & Model Size \\ \hline
	DS & 0.582 (0.044) & 28.5 \\
	TDS & 0.627 (0.051) & 9 \\
	Lasso & 0.590 (0.043) & 33 \\
	ALasso & 0.652 (0.051) & 8.5 \\
	Enet & 0.576 (0.040) & 67.5 \\
	CDS & 0.520 (0.025) & 9 \\ \hline
\end{tabular}
\end{table}

} 

\section{Discussion} \label{Sec5}

We have shown that the suggested constrained Dantzig selector can achieve convergence rates within a logarithmic factor of the sample size of the oracle rates in ultra-high dimensions under a fairly weak assumption on the signal strength. Our work provides a partial answer to an interesting question of whether convergence rates involving a logarithmic factor of the dimensionality are optimal for regularization methods in ultra-high dimensions. It would be interesting to investigate such a phenomenon for more general regularization methods.

Our formulation of the constrained Dantzig selector uses the $L_1$-norm of the parameter vector. A natural extension of the method is to exploit the weighted $L_1$-norm of the parameter to allow for different regularization on different covariates, as is in the adaptive Lasso (Zou, 2006). It would be interesting to investigate the behavior of these methods in more general model settings including generalized linear models and survival analysis. These problems are beyond the scope of the current paper and will be interesting topics for future research.\\





\appendix
\section*{Appendix: Proofs of main results}
\label{app:theorem}

\medskip

\noindent \textbf{Proof of Theorem \ref{T1}}

\medskip

\textbf{High probability event $\mathcal{E}$.} {Recall that $\lambda_0 = c_0 \sqrt{(\log n)/n}$ and $\lambda_1 = c_1 \sqrt{(\log p)/n}$.} All the results in Theorems \ref{T1} and \ref{T2} will be shown to hold on a key event
\begin{equation} \label{A001}
\mathcal{E} = \big\{\|n^{-1}\bX_1^T\bveps\|_{\infty} \leq \lambda_0 \ \text{ and } \ \|n^{-1}\bX_2^T\bveps\|_{\infty} \leq \lambda_1\big\},
\end{equation}
where $\bX_1$ is a submatrix of $\bX$ consisting of columns corresponding to $\supp(\bbeta_0)$ and $\bX_2$ consists of the remaining columns. Thus we will have the same probability bound in both theorems. The probability bound on the event $\mathcal{E}$ in (\ref{A001}) can be easily calculated, using the classical Gaussian tail probability bound (see, for example, Dudley, 1999) and the Bonferroni inequality, as
\begin{align} \label{A002}
\nonumber
\pr(\mathcal{E}) & \geq 1-\left\{\pr(\|n^{-1}\bX_1^T\bveps\|_{\infty} > \lambda_0)+\pr(\|n^{-1}\bX_2^T\bveps\|_{\infty} > \lambda_1)\right\} \\
\nonumber
& = 1-\big\{s (2/\pi)^{1/2} \sigma \lambda_0^{-1} n^{-1/2} e^{-\lambda_0^2 n/(2\sigma^2)}+
(p-s)(2/\pi)^{1/2} \sigma \lambda_1^{-1} n^{-1/2} e^{-\lambda_1^2 n/(2\sigma^2)}\big\} \\
& = 1- O\big\{s\ n^{-c_0^2/(2\sigma^2)}(\log n)^{-1/2} + (p-s) p^{-c_1^2/(2\sigma^2)}(\log p)^{-1/2}\big\},
\end{align}
{where the last equality follows from the definitions of $\lambda_0$ and $\lambda_1$.} Let $c = (c_0^2 \wedge c_1^2)/(2\sigma^2) - 1$ be a sufficiently large positive constant, since the two positive constants $c_0$ and $c_1$ are chosen large enough. Recall that $p$ is understood implicitly as $\max\{n, p\}$ throughout the paper. Thus it follows from (\ref{A002}), $s \leq n$, and $n \leq p$ that
\begin{equation} \label{A003}
\pr(\mathcal{E}) = 1-O\big(n^{-c}\big).
\end{equation}
From now on, we derive all the bounds on the event $\mathcal{E}$. In particular, in light of (\ref{A001}) and $\bbeta_0 \in \mathcal{B}_\lambda$ it is easy to verify that conditional on $\mathcal{E}$, the true regression coefficient vector $\bbeta_0$ satisfies the constrained Dantzig selector constraints; in other words, $\bbeta_0$ lies in the feasible set in (\ref{003}).

\medskip

\textbf{Nonasymptotic properties of $\hbbeta$.} We first make a simple observation on the constrained Dantzig selector $\hbbeta = (\hbeta_1, \ldots, \hbeta_p)\t$. Recall that without loss of generality, we assume $\supp(\bbeta_0) = \{1, \ldots, s\}$. Let $\bbeta_0 = (\bbeta_1\t, \textbf{0}\t)\t$ with each component of $\bbeta_1$ being nonzero, and $\hbbeta = (\hbbeta_1\t, \hbbeta_2\t)\t$ with $\hbbeta_1$ a subvector of $\hbbeta$ consisting of its first $s$ components. Denote by $\bdelta = (\bdelta_1\t, \bdelta_2\t)\t = \hbbeta - \bbeta_0$ the estimation error, where $\bdelta_1 = \hbbeta_1 - \bbeta_1$ and $\bdelta_2 = \hbbeta_2$. It follows from the global optimality of $\hbbeta$ that $\|\hbbeta_1\|_1 + \|\hbbeta_2\|_1 = \|\hbbeta\|_1 \leq \|\bbeta_0\|_1 = \|\bbeta_1\|_1$, which entails that
\begin{equation} \label{A004}
\|\bdelta_2\|_1 = \|\hbbeta_2\|_1 \leq \|\bbeta_1\|_1 - \|\hbbeta_1\|_1 \leq \|\hbbeta_1 - \bbeta_1\|_1 = \|\bdelta_1\|_1.
\end{equation}
We will see that this basic inequality $\|\bdelta_{2}\|_1 \leq \|\bdelta_{1}\|_1$ plays a key role in the technical derivations of both Theorem 1 and 2. Equipped with this inequality and conditional on $\mathcal{E}$, we are now able to start the derivation of all results in Theorem 1 as follows.

The main idea is to first prove its sparsity property which will be presented in the next paragraph. Then, with the control on the number of false positives and false negatives, we derive an upper bound for the $L_2$-estimation loss using the conclusion in Lemma 3.1 of Cand\`{e}s and Tao (2007). Results on other types of losses follow accordingly.

\medskip

(1) Sparsity.
Recall that under the assumption of Theorem 1, $\bbeta_0$ lies in its feasible set conditional on $\mathcal{E}$. {Since $\lambda_0 \leq \lambda_1$, by the definition of the constrained Dantzig selector, we have $|n^{-1} \bX\t (\by - \bX \hbbeta)| \preceq \lambda_1$, where $\preceq$ is understood as componentwise no larger than. Conditional on the event $\mathcal{E}$, substituting $\by$ by $\bX \bbeta_0 + \bveps$ and applying the triangle inequality yield
\begin{equation}\label{XXT}
\|n^{-1} \bX\t \bX \bdelta\|_\infty \leq 2\lambda_1.
\end{equation}
Furthermore, Lemma 3.1 in Cand\`{e}s and Tao (2007) still applies for $\bdelta$ as long as the uniform uncertainty principle condition (\ref{L005}) holds. Together with (\ref{A004}) and (\ref{XXT}), by applying the same argument as in the proof of Theorem 1.1 in Cand\`{e}s and Tao (2007), we obtain an $L_2$-estimation loss bound of $\|\bdelta\|_2 \leq (Cs)^{1/2}\lambda_1$ with $C = 4^2/(1 - \delta - \theta)^2$ some positive constant.}

{Since both the true regression coefficient vector $\bbeta_0$ and the constrained Dantzig selector $\hbbeta$ lie in the constrained parameter space $\mathcal{B}_\lambda$, the magnitude of any nonzero component in both $\bbeta_0$ and $\hbbeta$ is no smaller than $\lambda$. It follows that on the set of falsely discovered signs, the component of $\bdelta = \hbbeta - \bbeta_0$ will be no smaller than $\lambda$. Then making use of the obtained $L_2$-estimation loss bound $\|\bdelta\|_2 \leq (Cs)^{1/2}\lambda_1$, it is immediate that the number of falsely discovered signs is bounded from above by $C s(\lambda_1 / \lambda)^2$.} Under the assumption of $\lambda \geq C^{1/2}(1+ \lambda_1/\lambda_0)\lambda_1$, we have $C s(\lambda_1 / \lambda)^2 \leq s(1+\lambda_1/\lambda_0)^{-2} \leq s$.

\medskip

(2) $L_2$-estimation loss. We further exploit the technical tool of Lemma 3.1 in Cand\`{e}s and Tao (2007) to analyze the behavior of the estimation error $\bdelta = \hbbeta - \bbeta_0$. Let $\bdelta_1'$ be a subvector of $\bdelta_2$ consisting of the $s$ largest components in magnitude, $\bdelta_3 = (\bdelta_1\t, (\bdelta_1')\t)\t$, and $\bX_3$ a submatrix of $\bX$ consisting of columns corresponding to $\bdelta_3$. We emphasize that $\bdelta_3$ covers all nonzero components of $\bdelta$ since the number of falsely discovered signs is upper bounded by $s$, as showed in the previous paragraph. Therefore, $\|\bdelta_3\|_q = \|\bdelta\|_q$ for all $q > 0$. In view of the uniform uncertainty principle condition (\ref{L005}), an application of Lemma 3.1 in Cand\`{e}s and Tao (2007) results in
\begin{equation} \label{A005}
\|\bdelta_3\|_2 \leq (1-\delta)^{-1} \|n^{-1}\bX_3^T \bX\bdelta\|_2 + \theta(1-\delta)^{-1}s^{-1/2} \|\bdelta_2\|_1.
\end{equation}
On the other hand, from the basic inequality (\ref{A004}) it is easy to see that $s^{-1/2} \|\bdelta_2\|_1 \leq s^{-1/2} \|\bdelta_1\|_1 \leq \|\bdelta_3\|_2$. Substituting it into (\ref{A005}) leads to
\begin{equation} \label{A006}
\|\bdelta\|_2 = \|\bdelta_3\|_2 \leq (1-\delta-\theta)^{-1}\|n^{-1}\bX_3^T \bX\bdelta\|_2
\end{equation}
Hence, to develop a bound for the $L_2$-estimation loss it suffices to find an upper bound for $\|n^{-1} \bX_3^T \bX\bdelta\|_2$.

Denote by $A_1$, $A_2$, and $A_3$ the index sets of correctly selected variables, missed true variables, and falsely selected variables,  respectively. Let $A_{23} = A_{2}\cup A_{3}$. {Then we can obtain from the definition of the constrained Dantzig selector along with its thresholding feature that} $|n^{-1} \bX\t_{A_1} (\by - \bX \hbbeta)| \preceq \lambda_0$, $|n^{-1} \bX\t_{A_2} (\by - \bX \hbbeta)| \preceq \lambda_1$, and $|n^{-1} \bX\t_{A_3} (\by - \bX \hbbeta)| \preceq \lambda_0$. Conditional on $\mathcal{E}$, substituting $\by$ by $\bX \bbeta_0 + \bveps$ and applying the triangle inequality give
\begin{equation} \label{A007}
\|n^{-1} \bX\t_{A_1} \bX \bdelta\|_\infty \leq 2\lambda_0 \text{ \ and \ } \|n^{-1} \bX\t_{A_{23}} \bX \bdelta\|_\infty \leq \lambda_0 + \lambda_1.
\end{equation}
We now make use of the technical result on sparsity. Since $A_{23}$ denotes the index set of false positives and false negatives, its cardinality is also bounded by $C s(\lambda_1 / \lambda)^2$ with probability at least  $1-O(n^{-c})$. Therefore, by (\ref{A007}) we have
\begin{align*}
\|n^{-1} \bX_3^T \bX\bdelta\|_2^2 &\leq \|n^{-1}\bX_{A_1}^T \bX\bdelta\|_2^2 + \|n^{-1} \bX_{A_{23}}^T \bX\bdelta\|_2^2 \\ & \leq s\|n^{-1} \bX_{A_1}^T \bX\bdelta\|_\infty^2 + C s(\lambda_1 / \lambda)^2\|n^{-1} \bX\t_{A_{23}} \bX \bdelta\|_\infty^2 \\
& \leq 4s\lambda_0^2 + C s(\lambda_1 / \lambda)^2(\lambda_0 + \lambda_1)^2.
\end{align*}
Substituting this inequality into (\ref{A006}) yields
\[
\|\bdelta\|_2 \leq (1-\delta-\theta)^{-1}\big\{4s\lambda_0^2 + C s(\lambda_1 / \lambda)^2(\lambda_0 + \lambda_1)^2\big\}^{1/2}.
\]
Since we assume $\lambda \geq C^{1/2}(1+ \lambda_1/\lambda_0)\lambda_1$, it follows that $C (\lambda_1 / \lambda)^2(\lambda_0 + \lambda_1)^2 \leq \lambda_0^2$. Therefore, we conclude that $\|\delta\|_2 \leq (1-\delta-\theta)^{-1}(5s\lambda_0^2)^{1/2}$.

\medskip

(3) Other losses. Applying the basic inequality (\ref{A004}), we establish an upper bound for the $L_1$-estimation loss
\begin{align*}
\|\bdelta\|_1 & = \|\bdelta_1\|_1 + \|\bdelta_2\|_1 \leq 2 \|\bdelta_1\|_1 \leq 2 s^{1/2} \|\bdelta_1\|_2 \\
& \leq 2 s^{1/2} \|\bdelta\|_2 \leq 2(1-\delta-\theta)^{-1} s (5\lambda_0^2)^{1/2}.
\end{align*}
For the $L_\infty$-estimation loss, we additionally assume that $\lambda > (1-\delta-\theta)^{-1}(5s\lambda_0^2)$ which can lead to the sign consistency, $\sgn(\hbbeta) = \sgn(\bbeta_0)$, in view of the $L_2$-estimation loss inequality above. Therefore, by the constrained Dantzig selector constraints we have $\|n^{-1}\bX_1^T(\by - \bX_1 \hbbeta_1)\|_\infty \leq \lambda_0$ and thus $\|n^{-1} \bX_1^T(\bveps - \bX_1 \bdelta_1)\|_\infty \leq \lambda_0$. Then conditional on $\mathcal{E}$, it follows from the triangle inequality that $\|n^{-1} \bX_1^T \bX_1 \bdelta_1\|_\infty \leq 2 \lambda_0$. Hence, $\|\bdelta\|_\infty = \|\bdelta_1\|_\infty \leq 2 \|(n^{-1} \bX_1^T \bX_1)^{-1}\|_\infty\lambda_0$, which completes the proof.\\

\medskip

\noindent
\textbf{Proof of Theorem \ref{T2}}

\medskip

We continue to use the technical setup and notation introduced in the proof of Theorem \ref{T1}. Results are parallel to those in Theorem 1 but presented in the asymptotic manner. The similarity lies in the rationale of the proof as well. The key element is to derive the sparsity and then construct an inequality for $L_2$-estimation loss through the bridge $n^{-1}\| \bX\bdelta\|_2^2$. Inequalities for other types of losses are built upon this bound.

\medskip

(1) Sparsity. {Conditional on the event $\mathcal{E}$, we know that (\ref{A004}) and (\ref{XXT}) still hold by the definition of the constrained Dantzig selector. Moreover, Condition \ref{C1} is similar to the restricted eigenvalue assumption $\mathrm{RE}(s,m,1)$ in Bickel et al. (2009), except that $\mathrm{RE}(s,m,1)$ assumes the inequality holds for any subset with size no larger than $s$ to cover different possibilities of $\supp(\bbeta_0)$. Since we assume without loss of generality that $\supp(\bbeta_0) = \{1, \ldots, s\}$, an application of similar arguments as in the proof of Theorem 7.1 in Bickel et al. (2009) yields the $L_2$ oracle inequality} $\|\bdelta\|_2 \leq (C_m s)^{1/2}\lambda_1$ with $C_m$ some positive constant dependent on $m$. {Therefore, by the same arguments as in the proof of Theorem 1,} it can be shown that the number of falsely discovered signs is bounded from above by $C_m s(\lambda_1 / \lambda)^2$ and further by $s(1+\lambda_1/\lambda_0)^{-2}$ since $\lambda \geq C_m^{1/2}(1+ \lambda_1/\lambda_0)\lambda_1$. Next we will go through the proof of Theorem 7.1 in Bickel et al. (2009) in a more cautious manner and make some improvements in some steps with the aid of the obtained bound on the number of false positives and false negatives.

\medskip

(2) $L_2$-estimation loss. By Condition 1, we have a lower bound for $n^{-1}\|\bX\bdelta\|_2^2$. It is also natural to derive an upper bound for it and build an inequality related to the $L_2$-estimation loss. It follows from (\ref{A007}) that
\begin{equation} \label{A008}
\begin{aligned}
n^{-1}\|\bX\bdelta\|_2^2 & \leq \|n^{-1}\bX\t_{A_1}\bX \bdelta\|_\infty \|\bdelta_{A_1}\|_1 + \|n^{-1} \bX\t_{A_{23}} \bX \bdelta\|_\infty \|\bdelta_{A_{23}}\|_1 \\
& \leq 2\lambda_0\|\bdelta_{A_1}\|_1 + (\lambda_0 + \lambda_1)\|\bdelta_{A_{23}}\|_1.
\end{aligned}
\end{equation}
Since the cardinality of $A_{23}$ is bounded by $C_m s(\lambda_1 / \lambda)^2$, applying the Cauchy-Schwarz inequality to (\ref{A008}) leads to
$n^{-1}\|\bX\bdelta\|_2^2 \leq 2\lambda_0\|\bdelta_{A_1}\|_1 + (\lambda_0 + \lambda_1)\|\bdelta_{A_{23}}\|_1 \leq 2\lambda_0 s^{1/2} \|\bdelta_{A_1}\|_2 + C_m^{1/2}(\lambda_0 + \lambda_1)s^{1/2}\lambda_1 / \lambda\|\bdelta_{A_{23}}\|_2$.
This gives an upper bound for $n^{-1}\|\bX\bdelta\|_2^2$. Combining this with Condition 1 
results in
\begin{equation} \label{A009}
2^{-1}\kappa^2(\|\bdelta_{A_1}\|_2^2 + \|\bdelta_{A_{23}}\|_2^2) \leq 2\lambda_0 s^{1/2} \|\bdelta_{A_1}\|_2 + C_m^{1/2}(\lambda_0 + \lambda_1)s^{1/2}\lambda_1 / \lambda\|\bdelta_{A_{23}}\|_2.
\end{equation}
Consider (\ref{A009}) in a two-dimensional space with respect to $\|\bdelta_{A_1}\|_2$ and $\|\bdelta_{A_{23}}\|_2$. Then the quadratic inequality (\ref{A009}) defines a circular area centered at $(2\kappa^{-2}\lambda_0 s^{1/2},$ $\kappa^{-2}C_m^{1/2}(\lambda_0 + \lambda_1)s^{1/2}\lambda_1 / \lambda)$. The term $\|\bdelta_{A_1}\|_2^2 + \|\bdelta_{A_{23}}\|_2^2$ is nothing but the squared distance between the point in this circular area and the origin. One can easily identify the largest squared distance which is also the upper bound for the $L_2$-estimation loss
\[
\|\bdelta\|_2^2 = \|\bdelta_{A_1}\|_2^2 + \|\bdelta_{A_{23}}\|_2^2 \leq 4\kappa^{-4} \big\{ \big(2\lambda_0 s^{1/2}\big)^2 +  \big [C_m^{1/2}(\lambda_0 + \lambda_1)s^{1/2}\lambda_1 / \lambda \big]^2 \big\}.
\]
With the assumption of $\lambda \geq C_m^{1/2}(1+ \lambda_1/\lambda_0)\lambda_1$, we can show that $\big \{C_m^{1/2}(\lambda_0 + \lambda_1)s^{1/2}\lambda_1 / \lambda \big\}^2 \leq s \lambda_0^2$ and thus  $\|\bdelta\|_2 = O(\kappa^{-2}s^{1/2}\lambda_0)$. This bound has significance improvement with the factor $\log p$ reduced to $\log n$ in the ultra-high {dimensional} setting.

\medskip

(3) Other losses. With the $L_2$ oracle inequality at hand, one can derive the $L_1$ oracle inequality $\|\bdelta\|_1 \leq s^{1/2} \|\bdelta\|_2 = O(\kappa^{-2}s\lambda_0)$ in a straightforward manner. For the prediction loss, it follows from (\ref{A008}) that
\[
n^{-1}\|\bX\bdelta\|_2^2 \leq 2\lambda_0 s^{1/2} \|\bdelta_{A_1}\|_2 + C_m^{1/2}(\lambda_0 + \lambda_1)s^{1/2}\lambda_1 / \lambda\|\bdelta_{A_{23}}\|_2.
\]
Consider the problem of minimizing $2\lambda_0 s^{1/2} \|\bdelta_{A_1}\|_2 + C_m^{1/2}(\lambda_0 + \lambda_1)s^{1/2}\lambda_1 / \lambda\|\bdelta_{A_{23}}\|_2$ subject to (\ref{A009}). It is simply a two-dimensional linear optimization problem with a circular area as its feasible set. One can easily solve this minimization problem and obtain
\[ n^{-1}\|\bX\bdelta\|_2^2 \leq 2\kappa^{-2}\big[ \big(2\lambda_0 s^{1/2}\big)^2 +  \big \{C_m^{1/2}(\lambda_0 + \lambda_1)s^{1/2}\lambda_1 / \lambda \big\}^2 \big] = O(\kappa^{-2}s\lambda_0^2). \]
Finally, for the $L_\infty$ oracle inequality, it follows from similar arguments as in the proof of Theorem 1 that $\|\bdelta\|_\infty \leq 2 \|(n^{-1}\bX_1^T \bX_1)^{-1}\|_\infty\lambda_0 = O\{\|(n^{-1}\bX_1^T \bX_1)^{-1}\|_\infty\lambda_0\}$, which concludes the proof.\\

\medskip

\noindent
\textbf{Proof of Theorem \ref{T3}}

\medskip

Denote by $\hbbeta_{\mbox{global}}$ the global minimizer of (\ref{003}). Under the conditions of Theorem \ref{T2}, $\hbbeta_{\mbox{global}}$ enjoys the same oracle inequalities and properties as in Theorem \ref{T2} conditional on the event defined in (\ref{A001}). In particular, we have $FS(\hbbeta_{\mbox{global}}) = O(s)$. It follows that
\[ \|\hbbeta_{\mbox{global}}\|_0 \leq \|\hbbeta_0\|_0 + FS(\hbbeta_{\mbox{global}}) = O(s). \]
Let $\hbbeta$ be a computable local minimizer of (\ref{003}) produced by any algorithm satisfying $\|\hbbeta\|_0 \leq c_2 s$. Denote by $A = \supp(\hbbeta) \cup \supp(\hbbeta_{\mbox{global}})$. Then we have
\[
|A| \leq \|\hbbeta\|_0 + \|\hbbeta_{\mbox{global}}\|_0 = O(s) \leq c_3 s \] for some large enough positive constant $c_3$.

\medskip

We next analyze the difference between the two estimators, that is, $\bdelta = \hbbeta - \hbbeta_{\mbox{global}}$. Let $\bX_A$ be a submatrix of $\bX$ consisting of columns in $A$ and $\bdelta_A$ a subvector of $\bdelta$ consisting of components in $A$. Since $\|\bX\t(\by - \bX \hbbeta)\|_{\infty} = O(\lambda_1)$ and $\|\bX\t(\by - \bX \hbbeta_{\mbox{global}})\|_{\infty} = O(\lambda_1)$, we can show that
\[\|n^{-1}\bX_A\t \bX_A \bdelta_A\|_2 \leq |A|^{1/2}\|n^{-1}\bX_A\t \bX_A \bdelta_A\|_{\infty} = O(s^{1/2} \lambda_1).\]
By the assumption that $\min_{\|\bdelta\|_2 = 1,\ \|\bdelta\|_0 \leq c_3 s} n^{-1/2} \|\bX \bdelta\|_2 \geq \kappa_0$, the smallest singular value of $n^{-1/2} \bX_A$ is bounded from below by $\kappa_0$. Thus we have  $\|\bdelta\|_2 = \|\bdelta_A\|_2 = O(s^{1/2} \lambda_1)$. Together with the thresholding feature of the constrained Dantzig selector, it follows that the number of different indices between $\supp(\hbbeta)$ and $\supp(\hbbeta_{\mbox{global}})$ is bounded by $O\{s (\lambda_1/\lambda)^2\}$. 
This sparsity property is essential for our proof and similar sparsity results can be found in Theorem 2.

Based on the aforementioned sparsity property and by the facts that $\|\bX_{\widetilde{A}_1}\t \bX \bdelta\|_{\infty} \leq 2\lambda_0$ and $\|\bX_{\widetilde{A}_{23}}\t \bX \bdelta\|_{\infty} \leq \lambda_0 + \lambda_1$ with $\widetilde{A}_1 = \supp(\hbbeta) \cap \supp(\hbbeta_{\mbox{global}})$ and $\widetilde{A}_{23} = [\supp(\hbbeta) \setminus \supp(\hbbeta_{\mbox{global}})] \cup [\supp(\hbbeta_{\mbox{global}}) \setminus \supp(\hbbeta)]$, the same arguments as in the proof of Theorem \ref{T2} apply to show that  $\|\bdelta\|_2 = O(\kappa^{-2} s^{1/2} \lambda_0)$. It is clear that this bound is of the same order as that for the difference between $\hbbeta_{\mbox{global}}$ and $\bbeta_0$ in Theorem \ref{T2}. Thus $\hbbeta$ enjoys the same asymptotic bound on the $L_2$-estimation loss. Similarly, the asymptotic bounds for the other losses in Theorem \ref{T2} also apply to $\hbbeta$, since those inequalities are rooted on the bounds for the sparsity and  $L_2$-estimation loss. This completes the proof.\\


\vskip 0.2in
\bibliography{CDS}
\begin{description}
\item
Antoniadis, A., Fryzlewicz, P. and Letue, F. (2010). The Dantzig selector in Cox's proportional hazards model. \textit{Scand. J. Statist.} \textbf{37}, 531--552.

\item
Bickel, P. J., Ritov, Y. and Tsybakov, A. (2009). \newblock {Simultaneous analysis of Lasso and Dantzig selector}. \newblock \textit{Ann. Statist.} \textbf{37}, 1705--1732.

\item
Cand\`{e}s, E. J. and Tao, T. (2005). \newblock {Decoding by linear programming}. \newblock \textit{IEEE Trans. Info. Theory} \textbf{51}, 4203--4215.

\item
Cand\`{e}s, E. J. and Tao, T. (2007). \newblock {The Dantzig selector: statistical estimation when $p$ is much larger than $n$ (with Discussion)}. \newblock \textit{Ann. Statist.} \textbf{35}, 2313--2351.

\item
Cand\`{e}s, E. J., Wakin, M. B. and Boyd, S. (2008). \newblock {Enhancing sparsity by reweighted $l_1$ minimization}. \newblock \textit{J. Fourier Anal. Appl.} \textbf{14}, 877--905.

\item
Chen, K., Dong, H. and Chan, K. S. (2013). Reduced rank regression via adaptive nuclear norm penalization. \textit{Biometrika} \textbf{100}, 901--920.

\item
Donoho, D. L. (2006). \newblock {Compressed sensing}. \newblock \textit{IEEE Trans. Info. Theory} \textbf{52}, 1289--1306.

\item
Dudley, R. M. (1999). \newblock \textit{Uniform Central Limit Theorems}. \newblock Cambridge University Press.



\item
Fan, J. and Li, R. (2001). Variable selection via nonconcave penalized likelihood and its oracle properties. \textit{J. Amer. Statist. Assoc.} \textbf{96}, 1348--1360.

\item
Fan, J. and Lv, J. (2011). \newblock {Nonconcave penalized likelihood with NP-dimensionality}. \newblock \textit{IEEE Trans. Info. Theory} \textbf{57}, 5467--5484.

\item
Fan, J. and Peng, H. (2004). Nonconcave penalized likelihood with diverging number of parameters. \textit{Ann. Statist.} \textbf{32}, 928--961.

\item
Fan, Y. and Lv, J. (2013). \newblock {Asymptotic equivalence of regularization methods in thresholded parameter space}. \newblock \textit{J. Am. Statist. Assoc.} \textbf{108}, 1044--1061.

\item
Fan, Y. and Lv, J. (2014). \newblock {Asymptotic properties for combined $L_1$ and concave regularization}. \newblock \textit{Biometrika} \textbf{101}, 57--70.

\item
Fan, Y. and Tang, C. (2013). \newblock {Tuning parameter selection in high dimensional penalized likelihood}.\newblock \textit{J. R. Statist. Soc. Ser. {\rm B}} \textbf{75}, 531--552.

\item
van de Geer, S. (2008). \newblock {High-dimensional generalized linear models and the Lasso}. \newblock \textit{Ann. Statist.} \textbf{36}, 614--645.

\item
van de Geer, S., B\"{u}hlmann, P. and Zhou S. (2011). \newblock {The adaptive and the thresholded Lasso for potentially misspecified models (and a lower bound for the Lasso)}. \newblock \textit{Electronic Journal of Statistics} \textbf{5}, 688--749.

\item
Huang, J., Ma, S. and Zhang, C.-H. (2008). \newblock {Adaptive Lasso for sparse high-dimensional regression models}. \newblock \textit{Statistica Sinica} \textbf{18}, 1603--1618.

\item
Lan, H., Chen, M., Flowers, J., Yandell, B., Stapleton, D., Mata, C., Mui, E., Flowers, M., Schueler, K., Manly, K., Williams, R., Kendziorski, K. and Attie, A. D. (2006). \newblock {Combined expression trait correlations and expression quantitative trait locus mapping}. \newblock \textit{PLoS Genet.} \textbf{2}, e6.

\item
Lv, J. and Fan, Y. (2009). \newblock {A unified approach to model selection and sparse recovery using regularized least squares}. \newblock \textit{Ann. Statist.} \textbf{37}, 3498--3528.

\item
Negahban, S., Ravikumar, P., Wainwrigt, M. and Yu, B. (2012). \newblock {A unified framework for high-dimensional analysis of M-estimators with decomposable regularizers}. \newblock \textit{Statistical Science} \textbf{27}, 538--557.

\item
Raskutti, G., Wainwright, M. J. and Yu, B. (2011). \newblock {Minimax rates of convergence for high-dimensional regression under $\ell_q$-ball sparsity}. \newblock \textit{IEEE Trans. Inform. Th.} \textbf{57}, 6976--6994.

\item
Scheetz, T., Kim, K., Swiderski, R., Philp, A., Braun, T., Knudtson, K., Dorrance, A., DiBona, G., Huang, J., Casavant, T. and et al. (2006). \newblock {Regulation of gene expression in the mammalian eye and its relevance to eye disease}. \newblock \textit{Proc. Natl. Acad. Sci. USA} \textbf{103}, 14429--14434.

\item
Song, Q. and Liang, F. (2015). High dimensional variable selection with reciprocal $L_1$-regularization. \newblock \textit{J. Am. Statist. Assoc.}, in press.

\item
St\"{a}dler, N., B\"{u}hlmann, P. and van de Geer, S. (2010). \newblock {$\ell_1$-penalization in mixture regression models (with discussion)}. \newblock \textit{Test} \textbf{19}, 209--285.

\item
Tibshirani, R. J. (1996). \newblock {Regression shrinkage and selection via the Lasso}. \newblock \textit{J. R. Statist. Soc. Ser. {\rm B}} \textbf{58}, 267--288.

\item
Wainwright, M. J. (2009) \newblock {Sharp thresholds for noisy and high-dimensional recovery of sparsity using $\ell_1$-constrained quadratic programming (Lasso)}. \newblock \textit{IEEE Trans. Inform. Th.} \textbf{55}, 2183--2202.

\item
Xue, L. and Zou, H. (2011). \newblock {Sure independence screening and compressed random sensing}. \newblock \textit{Biometrika} \textbf{98}, 371--380.

\item
Zhang, C.-H. (2010) \newblock {Nearly unbiased variable selection under minimax concave penalty}. \newblock \textit{Ann. Statist.} \textbf{38}, 894--942.

\item
Zhang, T. (2011) \newblock {Adaptive forward-backward greedy algorithm for learning sparse representations}. \newblock \textit{IEEE Trans. Inform. Th.} \textbf{57}, 4689--4708.

\item
Zhao, P. and Yu, B. (2006). \newblock {On model selection consistency of Lasso}. \newblock \textit{J. Machine Learning Research} \textbf{7}, 2541--2567.

\item
Zheng, Z., Fan, Y. and Lv, J. (2014). \newblock {High-dimensional thresholded regression and shrinkage effect}. \newblock \textit{J. R. Statist. Soc. Ser. {\rm B}} \textbf{76}, 627--649.

\item
Zhou, T. and Tao, D. (2013). \newblock {Double shrinking sparse dimension reduction}. \newblock \textit{IEEE Transactions on Image Processing} \textbf{22}, 244--257.

\item
Zou, H. (2006). \newblock {The adaptive Lasso and its oracle properties}. \newblock \textit{J. Am. Statist. Assoc.} \textbf{101}, 1418--1429.

\item
Zou, H. and Hastie, T. J. (2005). \newblock {Regularization and variable selection via the elastic net}. \newblock \textit{J. R. Statist. Soc. Ser. {\rm B}} \textbf{67}, 301--320.

\end{description}

\end{document}